\documentclass[twocolumn,letterpaper,amsmath,amssymb,floatfix,aps,superscriptaddress]{revtex4} 

\usepackage{graphicx}
\usepackage{dcolumn}
\usepackage{bm}
\usepackage{epsfig,color}

\newcommand{\Av}[1]{{\mathbf #1}}

\def\ln{{\operatorname{ln}}}

\def\rmd{{\mathrm{d}}}
\def\rmi{{\mathrm{i}}}
\def\rme{{\mathrm{e}}}

\begin{document}

\title{Dressed Counterions: Strong Electrostatic Coupling in the Presence of Salt}

\author{Matej Kandu\v c}
\affiliation{Department of Theoretical Physics,
J. Stefan Institute, SI-1000 Ljubljana, Slovenia}

\author{Ali Naji}
\affiliation{Department of Physics and Astronomy, University of Sheffield, Sheffield S3 7RH, United Kingdom}

\author{Jan Forsman}
\affiliation{Department of Theoretical Chemistry, Lund University Chemical Center, P.O. Box 124, S-221 00 Lund, Sweden}

\author{Rudolf Podgornik}
\affiliation{Department of Theoretical Physics,
J. Stefan Institute, SI-1000 Ljubljana, Slovenia}
\affiliation{Institute of Biophysics, Medical Faculty and Department of Physics, Faculty of Mathematics and Physics,
University of Ljubljana, SI-1000 Ljubljana, Slovenia}

\begin{abstract}
We reformulate the theory of strong electrostatic coupling in order to describe an asymmetric electrolyte solution of monovalent salt ions and polyvalent counterions using field-theoretical techniques and Monte-Carlo simulations. The theory is based on an asymmetric treatment of the different components of the electrolyte solution. The weak coupling Debye-H\" uckel approach is used in order to describe the monovalent salt ions while 
a strong coupling approach is used to tackle the polyvalent counterions. This combined weak-strong coupling approach
effectively leads to dressed interactions between polyvalent counterions and thus directly affects the correlation attraction mediated by  polyvalent counterions between like-charged objects. The general theory is specifically applied to a system composed of two uniformly charged plane-parallel surfaces in the presence of salt and polyvalent counterions. In the strong coupling limit for polyvalent counterions the comparison with Monte-Carlo simulations shows good agreement for large enough values of the electrostatic coupling parameter. We delineate two limiting laws that in fact 
encompass all the Monte-Carlo data. 
\end{abstract}
\maketitle

\section{Introduction}

The properties of biological matter, soft matter and colloidal systems such as polyelectrolytes, charged membranes and charged micelles are in many important respects governed by the underlying properties of electrostatic interactions \cite{holm}. In these aqueous systems the charged macromolecular (macroion) surfaces are always surrounded by neutralizing counterions as well as salt ions from the electrolyte solution. In simple solutions of monovalent salt  (e.g., physiological solution of NaCl), the system is well described by the traditional Poisson-Boltzmann (PB) approach \cite{Andelman}. Being mean-field based, the PB theory is applicable only for sufficiently small macroion surface charge densities, low ion valencies, high medium dielectric constant and/or high temperatures. The limitations of this approach become practically important in highly charged systems when ion-ion correlation effects start to govern the electrostatic properties of the system \cite{holm,hoda,Naji,shklovskii,Levin, Rouzina,Netz,messina}. In some sense the accuracy of the PB approach can be systematically improved by perturbative corrections in the ionic fluctuations and correlations \cite{podgornik2,attard,podgornik,fluctuations,david-ron,asim,ziherl} along the lines of the standard approach used in the mean-field context. But the validity of this approach is severely limited since the perturbative (loop) expansion is weakly convergent  \cite{Netz,hoda,Naji} and  higher-order corrections beyond the first-loop correction (Gaussian fluctuations) are 
very complicated \cite{david-ron,podgornikparsegianPRL} and in many cases intractable to evaluate. 
Thus, in many cases such perturbative corrections offer only an insignificant improvement over a PB approximation \cite{hoda,Naji} and can therefore not be used to predict phenomena 
such as like-charge attraction that differ qualitatively from the mean-field predictions \cite{original_sims}.

An alternative approach has been pioneered by several workers \cite{Rouzina,shklovskii,Levin,Netz} using various analytical methods that allow for non-perturbative treatment 
of correlations in highly charged systems.  A deeper insight into the electrostatic properties of Coulomb fluids led to the conclusion that their behavior 
lies between two different limiting cases: 
the {\em weak coupling} (WC) limit, governed by the PB theory, and the {\em strong coupling} (SC) limit, governed by 
the so-called strong coupling theory  \cite{hoda,Naji,Netz}. In fact, 
these two limits can be established \cite{Netz} as two exact {\em limiting laws} from a single  general theory  and thus allow for an exact treatment of charged systems  at two disjoint limiting conditions. The parameter space in between can be analyzed by approximate methods \cite{Forsman04,intermediate_regime,Weeks} that 
 cross over between the two limits and is mostly accessible solely via computer simulations \cite{hoda,Naji,Netz,arnold,asim,Weeks,Jho1,original_sims,Forsman04,trulsson,jho-prl,Naji_CCT}. 
 Exact solutions  for the whole range of coupling parameters are available only in one
dimension \cite{exact}. The WC-SC paradigm has been tested extensively \cite{hoda,Naji,Netz,arnold,asim,Forsman04,intermediate_regime,Weeks,Jho1,jho-prl,Naji_CCT,exact} and was found to describe computer simulations quantitatively correctly in the respective regimes of validity, thus providing a unifying view of the behavior of Coulomb fluids. 

However, the SC theory was so far designed exclusively to treat counterion-only systems,  i.e., a Coulomb fluid composed of only counterions in the absence of any salt ions \cite{Netz}. Though an approximation of this type can be used to describe situations where a large amount of polyvalent counterions dominate the electrostatic behavior, it is in general unrealistic and has to be amended in order to deal with the complexity of real systems that always contain variable amounts of simple salt  \cite{Israelachvili}. A common situation would be a system composed of fixed surface charges with polyvalent counterions bathed in a solution of simple monovalent salt \cite{rau-1,rau-2}. 
This leads to a difficult problem of asymmetric aqueous electrolytes where different components of the Coulomb fluid are differently coupled to local electrostatic fields \cite{olli}. Some components, such as polyvalent counterions, are coupled strongly, whereas some, such as monovalent salt ions, are coupled weakly. In systems of this type, no single approximation scheme that would treat all the charged components on the same level would be expected to work. Whereas the SC framework would certainly work for the polyvalent counterions, it would fail 
for the monovalent salt. The converse is true for the WC framework. In providing a consistent theoretical framework to describe highly asymmetric electrolytes, one is thus faced with a conundrum since no single approximation scheme appears to be valid in any range of coupling parameters. In what follows we will build a theoretical framework that will allow us to {\em selectively} use different approximation schemes for different components of the asymmetric Coulomb fluid. This combined WC-SC approach appears to bring forth all the salient features of these asymmetric systems at high electrostatic couplings that can be observed in Monte-Carlo (MC) simulations. 

Both the WC and the SC limits  of a system composed of fixed surface charges with intervening mobile counterions can be obtained based on a functional integral or field-theoretic representation  \cite{podgornik} of the grand canonical partition function. The functional integral representation of the Coulomb fluid partition function was pioneered by Edwards and Lenard in the 1960s \cite{Edwards} and has found many applications in the modern theory of Coulomb fluids. It was shown by Netz \cite{Netz} that both limits can be obtained by systematic expansion with respect to a
single electrostatic coupling parameter, which is connected closely to the plasma parameter and quantifies the relative importance of interactions between the mobile charges {\em versus} their interaction with a fixed external charge distribution of opposite sign. It can be introduced in the following way: the typical distance at which two unit charges $e_0$ interact with thermal energy $k_{\mathrm{B}}T$ in a medium of dielectric constant  $\varepsilon$ is known as the {\em Bjerrum length}, $\ell_{\mathrm{B}}=e_0^2/(4\pi\varepsilon\varepsilon_0 k_{\mathrm{B}}T)$ (in water at room temperature, the Bjerrum length is $\ell_{\mathrm{B}}\approx 0.7$ nm). In that respect the Bjerrum length for two counterions with valencies $+q$ is then $q^2 \ell_{\mathrm{B}}$. Similarly, the typical distance at which a counterion interacts with a charged surface (with surface charge density of $-\sigma$) with an energy equal to $k_{\mathrm{B}}T$ is given by the {\em Gouy-Chapman} length, defined as  $\mu_{\mathrm{GC}}=e_0/(2\pi q\ell_{\mathrm{B}}\sigma)$. A competition between counterion-counterion and counterion-surface interactions can be quantitatively measured by the ratio of these two characteristic length scales, i.e.
\begin{equation}
\Xi=q^2 \ell_{\mathrm{B}}/\mu_{\mathrm{GC}}=2\pi q^3 \ell_{\mathrm{B}}^2\sigma/e_0,
\label{xidef}
\end{equation}
which is known as the (Netz-Moreira) electrostatic coupling parameter \cite{Netz}.

The WC regime is characterized by a small coupling parameter ${\Xi}\ll 1$. It is governed by the mean-field  PB theory (obtained in the exact limit of $\Xi\rightarrow 0$) and
weak Gaussian fluctuations around the PB solution. 
This regime is adequate for low valency counterions and/or weakly charged surfaces and represents the physical situation where 
the width of the counterion layer $\mu_{\mathrm{GC}}$ close to the oppositely charged surface is much larger than the separation between two neighboring counterions
in the layer. Thus the counterion layer behaves essentially as a three-dimensional gas and a collective mean-field approach of the PB type is completely 
justified. For low electrostatic potentials the PB equation may be further simplified into the linear form known as the Debye-H\"uckel (DH) equation. 

On the other hand, in the SC regime ${\Xi}\gg 1$ (appropriate for high valency counterions and/or highly charged surfaces), the mean distance between counterions, $a_\bot \sim \sqrt{qe_0/\sigma}$ 
(as set by the local electroneutrality condition), is much larger than the counterion layer width (i.e., $a_\bot/\mu_{\mathrm{GC}}\sim \sqrt{\Xi}\gg 1$) \cite{hoda,Naji}, indicating that the counterions are highly localized laterally and form a strongly correlated quasi-two-dimensional layer right next to an oppositely charged surface. In this case, the WC approach breaks down due to strong counterion-surface and counterion-counterion interactions. Since each counterion is isolated laterally in a large {\em correlation hole} and can move almost independently from all the others along the direction perpendicular to the surface, the collective many-body effects diminish and on the leading order, in the limit $\Xi\rightarrow \infty$, the behavior of the system can be described {\em exactly} by a single-particle SC theory which includes only the contribution from counterion-surface interactions \cite{Netz}. 

The range of validity of the aforementioned WC and SC theories (when applied to systems with finite values of the coupling parameter) has been explored thoroughly in Refs. \cite{hoda,Naji,Netz}. Interestingly, although both these theories are obtained as limiting results, they can still be applied to charged systems in a wide range of realistic system parameters as confirmed by extensive numerical simulations \cite{hoda,Naji,Netz,arnold,asim, Forsman04,Weeks,Jho1,jho-prl,Naji_CCT,exact}. 

The above framework was introduced originally for the case where the mobile charges belong to a single species of counterions \cite{Netz}. In what follows we will attempt to formulate a systematic description for an asymmetric system containing polyvalent counterions and monovalent added salt and delineate its various limiting behaviors in a unifying theoretical framework. In the SC limit,  we shall use the standard SC procedure which is based on a virial and $1/\Xi$ expansion of the partition function \cite{Netz,Naji,hoda}, whereas in the WC limit, we shall perform a saddle point evaluation of the functional integral representation of the partition function \cite{podgornik,Netz}. The main approximation that we introduce in the SC limit is to treat the salt ions as weakly coupled to all the charges in the system. Their presence in the system thus has a simple consequence, namely, it renormalizes the Coulomb kernel into a Debye-H\"uckel kernel. The counterions can be treated, however,  either as weakly or strongly coupled depending on, e.g., their charge valency. As we show, this approximation makes sense only when one deals with a highly asymmetric system such as the one considered here. 

In the developments as detailed in what follows, the WC and the SC approaches do not carry an equal weight. In
fact we include the WC approximation in our discussion only for reasons of completeness since it reduces 
to well-known and  established results in the framework of the PB theory. The main thrust of this paper is 
thus devoted almost exclusively to the SC limit in the case of highly asymmetric systems that have thus far 
eluded a consistent statistical mechanical formulation.

\section{Model}
\label{sec:model}

We set up a simple model system that would exhibit the most important features of a typical asymmetric Coulomb fluid. It is composed of a 
fixed charge distribution  (macroions), $\rho_0(\Av r)$, 
and polyvalent counterions of charge valency $+q$ in a bathing
salt solution. For simplicity we consider a simple symmetric salt
with monovalent co- and counterions (the results can be
generalized immediately to any number of salt species as long as
salt ions stay weakly coupled to other charge species as will be 
made clear later). 
We shall later on apply our theory to the specific case 
where fixed charges are present on two plane-parallel surfaces (of infinite area $S$) 
located at positions $z=\pm a$, each carrying a uniform surface charge density of $-\sigma$, i.e., in this case
\begin{equation}
\rho_0(\Av r) = -\sigma[\delta(z-a)+\delta(z+a)]. 
\label{rho_0_plan}
\end{equation}

\begin{figure}[t]
\centerline{\psfig{figure=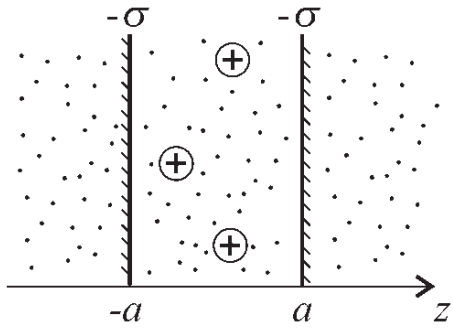,width=6cm}}
\caption{Schematic view of two negatively charged surfaces with uniform surface charge density $-\sigma$ and polyvalent counterions in between. Salt ions are present everywhere is space. }
\label{geom}
\end{figure}

The polyvalent counterions are assumed to be confined to the region
in between the charged surfaces, whereas the salt ions are assumed to be present in all regions in space (Fig.~\ref{geom}). In what follows we may refer to the $q$ valency (polyvalent)  counterions simply as  ``counterions"; they should not
be confused with salt cations which we shall refer to as  ``salt counterions" or  ``cations".
We shall neglect the effects of image charges arising from dielectric discontinuity at the bounding surfaces. 
The effects due to an inhomogeneous distribution of salt in the system as well as the image charges  can be included in a straightforward manner
within our formalism \cite{rudiali,kanduc,jho-prl}  and will be considered elsewhere.

So far we have invoked no constraints on the number of polyvalent counterions, $N$, between the surfaces. We will consider two different ensembles for   counterions, namely,  the canonical ensemble, where the
number of polyvalent  counterions  is fixed in the system, and the grand-canonical ensemble, where polyvalent counterions can be exchanged between the system and the bulk ionic  ``reservoir" and
$N$ is thus determined by assuming chemical equilibrium with the bulk. Note that the salt ions are always treated within the grand-canonical ensemble,
i.e., they are assumed to be in equilibrium with a bulk reservoir. We also introduce a dimensionless counterion fraction parameter $\eta$ as the amount of polyvalent  counterions between the surfaces relative to the surface charge,
\begin{equation}
\eta=\frac{Ne_0 q}{2\sigma S}.
\label{eta}
\end{equation}
The case $\eta=0$ represents a system with salt only, and $\eta=1$ is the case when the total charge due to counterions exactly compensates the surface charge. Note that in the canonical case the parameter $\eta$ is fixed, whereas it is a function of inter-surface distance in the grand-canonical case. In a  system with counterions only the global electroneutrality condition stipulates that $\eta=1$. This no longer needs to be the case and $\eta$ can take any non-negative value when salt ions are present, which--due to their screening properties--turn the long range Coulomb potential into a short range DH potential (see below) and can thus take care of the electroneutrality condition themselves.

\section{General formalism}

The total electrostatic interaction energy of  a given configuration of the system can be written as
\begin{equation}
W=\frac 12 \int\rho(\Av r)u(\Av r, \Av r')\rho(\Av r')\rmd \Av r\,\rmd\Av r',
\end{equation}
where $u(\Av r, \Av r')$ is the Coulomb kernel given by 
\begin{equation}
u(\Av r, \Av r')=\frac{1}{4\pi\varepsilon\varepsilon_0\vert\Av r-\Av r'\vert},
\end{equation}
and $\rho(\Av r)$ is  the total charge density comprising both the fixed charge distribution as well as the charge distribution of mobile charges, 
\begin{eqnarray}
\rho(\Av r)&=&\rho_0(\Av r)+\sum_i e_0 q\delta(\Av R_i^c\!\!-\Av r)+\nonumber\\
	&&+\sum_i e_0\delta(\Av R_i^+\!\!-\Av r)-\sum_i e_0\delta(\Av R_i^-\!\!-\Av r). 
\end{eqnarray}
Here $\Av R_i^{c}$, $\Av R_i^{+}$ and $\Av R_i^{-}$ are the positions of the polyvalent counterions, monovalent cations (salt counterions) and monovalent anions (salt coions), respectively.  
We then follow the standard procedure by introducing a fluctuating local potential, $\phi$, via the Hubbard-Stratonovich transformation, which leads to the following 
exact functional integral representation for the grand-canonical partition function \cite{podgornik,Netz}
\begin{equation}
{\cal Z}_G=\int{\cal D}\phi\,\rme^{-\beta H[\phi]},
\label{partfun}
\end{equation}
where $\beta=1/(k_{\mathrm{B}}T)$ and the field-functional Hamiltonian reads
\begin{eqnarray}
H[\phi]&=&\frac 12\int\!\!\!\!\int\phi(\Av r)u^{-1}(\Av r,\Av r')\phi(\Av r')\rmd\Av r\rmd\Av r'  + \rmi\int\rho_0(\Av r)\phi(\Av r)\rmd \Av r\nonumber\\
&-&\frac{\Lambda_c}{\beta}\int\rme^{- \rmi\beta e_0 q\phi(\Av r)}\Omega(\Av r)\rmd \Av r  \nonumber\\
&-& \frac{\Lambda_+}{\beta}\int\rme^{-\rmi\beta e_0\phi(\Av r)}\rmd \Av r
- \frac{\Lambda_-}{\beta}\int\rme^{\rmi\beta e_0\phi(\Av r)}\rmd \Av r
\label{hamiltonian}
\end{eqnarray}
Here $u^{-1}(\Av r,\Av r')$ is the inverse Coulomb kernel, 
\begin{equation}
u^{-1}(\Av r,\Av r')=-\varepsilon\varepsilon_0 \nabla^2 \delta(\Av r-\Av r'),
\end{equation}
and $\Lambda_c$ and $\Lambda_{\pm}$ represent the fugacities of polyvalent counterions and salt ions. The geometry function $\Omega(\Av r)$ specifies the
region accessible to polyvalent counterions; for the plane-parallel system in Fig.~\ref{geom}, we have $\Omega(\Av r)=1$ for $|z|<a$ and zero otherwise. 
The effective Hamiltonian (\ref{hamiltonian}) involves nonlinear terms 
that result from integrating out the microscopic degrees of freedom due to 
counterions and salt ions and elude an exact analytical treatment. We shall thus concentrate on the asymptotic behavior
of the system in the WC and SC limit where we show that the problem can be handled analytically 
for an asymmetric system as considered in this work. 

In the WC limit (for both the counterions as well as the salt ions), 
the above functional integral is dominated by the saddle point  of the Hamiltonian $H[\phi]$ \cite{Netz} 
which leads to the standard PB theory. In
an asymmetric system with large $q $ the monovalent salt ions can be treated in the linearized DH limit as we 
shall  discuss in Section \ref{sec:PB} below.

In the SC limit, the WC approximation breaks down and one should apply a different approach that we 
shall refer to as the {\em SC dressed counterion theory}. In this approach, due to the high asymmetry for 
large $q$, the monovalent salt ions can be treated again 
as weakly coupled on the DH level even though the counterions are treated as strongly coupled. 
This situation will be considered in Sections \ref{sec:SC-formalism}-\ref{sec:SC-grandcanon}. 
It will be shown that these asymptotic theories constitute two limiting laws that in fact bracket all the Monte-Carlo data.

\section{Weak coupling limit: Mean-field PB theory}
\label{sec:PB}

As already stated in the Introduction we include the discussion of the WC limit only for reasons of 
completeness. 
The WC limit does not venture outside the standard PB theory except marginally in the fact that 
we derive a limiting form of the standard  PB equation valid in the case of a highly asymmetric ionic 
system when salt ions are monovalent but counterions have a larger valency $q$, adjusting other 
physical parameters of the theory (such as the surface charge density) 
in such a way that the coupling parameter for salt ions as well as counterions remains by assumption small. 
The thus constructed PB theory   
in fact represents the {\em formal} weakly coupled limit of the strongly coupled theory of dressed counterions
to be introduced later. 

As is well known from the previous work on the counterion-only systems,  the WC limit corresponds to the saddle-point evaluation of the partition function in the limit of a vanishing coupling parameter \cite{Netz,hoda,Naji}, i.e. via the condition 
\begin{equation}
\delta H[\phi]/\delta \phi|_{\phi_{\mathrm{SP}}}=0.
\end{equation}  
This subsequently leads to the standard PB equation determining the real-valued mean-field electrostatic potential $\psi={\mathrm{i}} \phi_{\mathrm{SP}}$, that is 
\begin{eqnarray}
-\varepsilon\varepsilon_0 \nabla^2\psi&=&e_0 q \Lambda_c \Omega(\Av r) \rme^{-\beta e_0 q \psi} + \nonumber\\
&& + \,e_0\Lambda_+ \rme^{-\beta e_0 \psi}-e_0\Lambda_- \rme^{\beta e_0 \psi}  = \nonumber\\
&=& e_0q \,n_c(\Av r) + e_0 n_+(\Av r) - e_0 n_-(\Av r).
\label{PB_0}
\end{eqnarray}
Note that the terms on the right hand side represent the PB charge densities due to valency $q$ counterions, $e_0q \,n_c(\Av r)$, and the monovalent salt ions, $\pm e_0 n_\pm(\Av r)$, respectively. 

The boundary conditions at both surfaces require the continuity of the potential $\psi$ and jump in its first derivative due to the surface charge,
\begin{equation}
\psi'(\pm a)_\textrm{in}-\psi'(\pm a)_\textrm{out}=\mp\frac{\sigma}{\varepsilon\varepsilon_0}.
\end{equation}
There are two ways to proceed now depending on the thermodynamic ensemble chosen for the counterions. As noted before, the salt ions are 
assumed to be exchangeable with the bulk reservoir and are thus treated always
within a grand-canonical description. But the number of counterions may be taken to be fixed or determined via an equilibrium with 
the bulk. 

\subsection{Canonical ensemble}
\label{subsec:WC_canon}

In the canonical ensemble the counterions are not in equilibrium with a bulk reservoir. Their number in the slit is fixed and given by
\begin{equation}
\Lambda_c  \int \Omega(\Av r) \rme^{-\beta e_0 q\psi} \rmd \Av r = N.
\label{fixedN}
\end{equation}
As already stated, we will express the number of counterions through the parameter $\eta$, Eq. (\ref{eta}).
On the other hand,  the salt ions are in equilibrium with a bulk reservoir that contains equal amounts of both positive and negative ions, which implies that they have equal 
fugacities $\Lambda_+=\Lambda_-\equiv n_0$, giving  the  DH screening parameter $\kappa$ (inverse ``screening length") as
\begin{equation}
\kappa^2=8\pi \ell_{\mathrm{B}} n_0.
\end{equation}

At this stage it is convenient to introduce dimensionless quantities, namely
\begin{equation}
{\bf\tilde r}=\Av r/\mu_{\mathrm{GC}},\quad\tilde \kappa=\kappa\,\mu_{\mathrm{GC}},\quad u=\beta e_0 q \psi, 
\label{dimless}
\end{equation}
where $\mu_{\mathrm{GC}}$ is the Gouy-Chapman length as defined before and $u$ is the  dimensionless potential.

The dimensionless PB equation now assumes the form 
\begin{equation}
u''=\kappa^2 q\sinh\frac{u}{q}+C\,\rme^{-u}.
\end{equation}
In the case of a highly asymmetric ionic system with $q \gg 1$, one can linearize the salt ions term in the PB equation, leaving the counterion term in the non-linear form. This assumption is valid if the dimensionless potential $u/q$ is small enough, leading to the PB equation in the space between the surfaces $\vert z\vert<a$ of the form
\begin{equation}
u''=\kappa^2 u-C\rme^{-u},
\label{canon-PB}
\end{equation}
which is independent of the valency $q$. The constant $C$ can be evaluated when one stipulates the fixed amount of counterions, eq.~(\ref{fixedN}), which gives $C\int_{-a}^a \exp(-u)\,\rmd z=4\eta$. In the same limit the PB equation outside the surfaces, $\vert z\vert>a$, has the form $u''=\kappa^2 u,\label{PBout}$ which yields the well-known solutions $u(z)=u_0\exp(\pm \kappa z)$.

\begin{figure}[t]\begin{center}
	\begin{minipage}[b]{0.36\textwidth}\begin{center}
		\includegraphics[width=\textwidth]{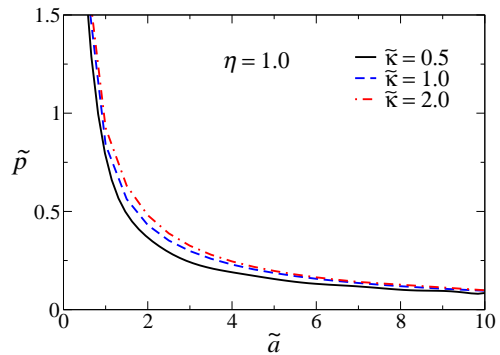}
	\end{center}\end{minipage} \hskip0.25cm
\caption{(Color online) Rescaled canonical PB pressure (\ref{eq:PB_DH_C})
as a function of the  rescaled inter-surface 
half-distance $\tilde a$ for $\eta=1$ and different values of the screening parameter $\tilde\kappa$ as shown on the graph.  } 
\label{fig_PBC}
\end{center}\end{figure}

The interaction pressure, $p$, between the bounding surfaces is given by the difference of the ion concentrations at the mid-plane ($z=0$), where the mean electric field vanishes, and the bulk concentration, i.e., $\beta p=n_+(0)+n_-(0)+n_c(0)-2n_0$, which leads to the dimensionless expression 
\begin{equation}
\tilde p = \frac 14 \tilde\kappa^2 u^2(0)+\frac{1}{2}C \rme^{-u(0)},
\label{eq:PB_DH_C}
\end{equation}
where we have defined the dimensionless pressure 
\begin{equation}
\quad\tilde p =\frac{2\varepsilon\varepsilon_0}{\sigma^2}\,p.
\end{equation}
As evident from the above equation, the pressure can never be negative, i.e., the effective interaction 
between two like-charged surfaces is always repulsive within this type of mean-field approach (this statement is true also within the full nonlinear theory) \cite{PBrepulsive}. 
The canonical PB equation  (\ref{canon-PB})
can be solved numerically for the dimensionless potential $u$. The results for the PB pressure  
as a function of the inter-surface separation are shown in Fig.~\ref{fig_PBC}  for $\eta=1$ and several different values of $\tilde \kappa$. The pressure 
depends  only weakly on the 
screening parameter due to dominant osmotic contributions from counterions.

\subsection{Grand-canonical ensemble}
\label{subsec:WC_grandcanon}
 
In the grand canonical ensemble the counterions too are in equilibrium with the bulk. In what follows the bulk concentration of counterions $c_0$ will be expressed by the parameter 
\begin{equation}
\chi^2=8\pi q^2 \ell_{\mathrm{B}} c_0. 
\end{equation}
The amount of positive $n_0^{*(+)}$ and negative $n_0^{*(-)}$ salt ions in the bulk containing also counterions is not equivalent. Here the electroneutrality condition demands that $n_0^{*(+)}+qc_0=n_0^{*(-)}$.

We allow bulk counterions to enter only the space between the surfaces (i.e., inside the slit $\vert z\vert<a$), but they are not allowed to pass through the bounding surfaces
and enter the ``outside"  region $\vert z\vert>a$. In contrast, salt ions are in equilibrium both with the bulk as well as the region $\vert z\vert>a$. 
In this situation the Donnan equilibrium is established between the bulk reservoir 
and the  ``outside" region, giving therefore 
\begin{eqnarray}
n_0^{*(\pm)}&=&n_0\,\rme^{\mp\beta e_0\psi_D},
\end{eqnarray}
where $n_0$ is the salt concentration infinitely far from the charged surfaces in the ``outside"  region and $\psi_D$ is the corresponding Donnan potential (i.e., the 
 potential difference between the bulk  and the  region  $\vert z\vert>a$ where the potential is in fact assumed to be zero  infinitely far from the charged surfaces). 
From the above equations we obtain  
\begin{equation}
\sinh\frac{u_D}{q} = \frac{qc_0}{2n_0} = \frac{\chi^2}{2\kappa^2q}
\label{eq:u_D_def}
\end{equation}
for  the dimensionless Donnan potential $ u_D=\beta e_0 q \psi_D$, which, 
   to the first order (appropriate for the linearized theory), gives $u_D = {\chi^2}/{2\kappa^2}$. Note that we have defined the screening parameter $\kappa$ in terms of the concentrations in the  ``outside" region as $\kappa^2=8\pi \ell_{\mathrm{B}} n_0$, which is identical to the one used in the canonical case. We also have $\Lambda_+ = \Lambda_- =n_0$ and $\Lambda_c=c_0 e^{u_D}$.

\begin{figure}[t]\begin{center}
	\begin{minipage}[b]{0.34\textwidth}\begin{center}
		\includegraphics[width=\textwidth]{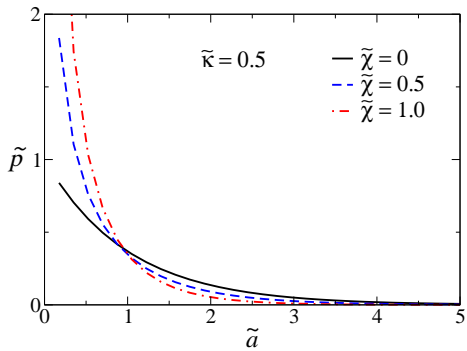}
	\end{center}\end{minipage} \hskip0.25cm
\caption{(Color online) Rescaled grand-canonical PB pressure (\ref{eq:P_PB_GC})
as a function of the rescaled inter-surface 
half-distance $\tilde a$ for $\tilde\kappa=0.5$ and different values of the parameter $\tilde\chi$ as shown on the graph.   } 
\label{fig_PBGC}
\end{center}\end{figure}

The grand-canonical PB equation in the slit region takes  the following form for the dimensionless potential, 
\begin{eqnarray}
u''=\kappa^2 q\sinh\frac{u}{q}-\tfrac 12\chi^2\,\rme^{-(u-u_D)}.
\label{eq:PB_GC_full}
\end{eqnarray}
After linearization of the salt part (first term on the right hand side of Eq. (\ref{eq:PB_GC_full})) in the case of a highly asymmetric system with large $q$, the PB equation assumes the form
\begin{equation}
u''=\kappa^2 u-\tfrac 12\chi^2\rme^{-(u-u_D)}.
\label{gcanon-PB}
\end{equation}
In the  region $\vert z\vert>a$,  the PB equation has the same form and solutions as in the canonical case, see Section \ref{PBout}. We furthermore stipulate the same boundary conditions as in the canonical case. The inter-surface pressure
is again given by the difference of the ion concentrations at the mid-plane ($z = 0$) and in the bulk so that $\beta p = n_+(0) + n_-(0) + n_c(0) - n_0^{*(+)} - n_0^{*(-)} - c_0$. The dimensionless pressure then assumes the form
\begin{equation}
\tilde p=\frac 14 \tilde\kappa^2(u^2(0)-u_D^2)+\frac 14 \tilde \chi^2\bigl[\rme^{-(u(0)-u_D)}-1\bigr],
\label{eq:P_PB_GC}
\end{equation}
where we have  introduced the dimensionless parameter 
\begin{equation}
\tilde \chi = \chi \,\mu_{\mathrm{GC}}. 
\end{equation}
One can again show that the grand-canonical pressure is always positive within the PB theory, just as in the canonical
 case \cite{PBrepulsive}. 
The PB equation in the grand-canonical ensemble, Eq. (\ref{gcanon-PB}), 
is solved numerically. The results for the PB pressure are shown in Fig.~\ref{fig_PBGC} for $\tilde \kappa=0.5$ and several different values of $\tilde \chi$. 
Note that the grand-canonical pressure remains finite also in the limit of vanishing inter-surface separation as in this case
all the counterions can be expelled from the inter-surface region.


\section{SC Dressed counterion theory: Formalism}
\label{sec:SC-formalism}

We now turn our attention to the main part of this paper, namely the case with strongly coupled counterions.
Proceeding from the Hamiltonian (\ref{hamiltonian}), we introduce the following approximation: we 
assume that only the polyvalent counterions are strongly coupled, whereas the monovalent salt ions are weakly coupled to other 
charges in the system. Since salt ions are in equilibrium with a bulk reservoir, the fugacities of salt counterions and coions are the same, $\Lambda_+=\Lambda_-\equiv n_0$, and have exactly the same meaning as in the previous Section.


The salt terms (the last two terms) in Eq. (\ref{hamiltonian}) can be expressed as
$\cos\,\beta e_0\phi(\Av r)$ and for a highly asymmetric system they can be expanded up to the quadratic 
order in the fluctuating potential. 
Thus up to an irrelevant constant we remain with
\begin{eqnarray}
H_{\mathrm{eff}}[\phi]&=&\frac 12\int\!\!\!\!\int\phi(\Av r)u_{\rm DH}^{-1}(\Av r,\Av r')\phi(\Av r')\rmd\Av r\rmd\Av r' + \rmi\int\rho_0(\Av r)\phi(\Av r)\rmd \Av r\nonumber\\
&-&\frac{\Lambda_c}{\beta}\int\rme^{- \rmi\beta e_0 q\phi(\Av r)} \Omega(\Av r)\rmd \Av r.
\label{hamiltonian-renorm}
\end{eqnarray}
This procedure therefore yields an effective Hamiltonian for a ``counterion-only" system but with the proviso that the inverse Coulomb kernel is replaced by the standard inverse Debye-H\" uckel kernel 
\begin{equation}
u_{\rm DH}^{-1}(\Av r,\Av r')=-\varepsilon\varepsilon_0(\nabla^2-\kappa^2)\delta(\Av r-\Av r'),
\label{DHinv}
\end{equation}
where $\kappa^2 = 8\pi \ell_{\mathrm{B}}n_0$.
We have thus effectively integrated out the salt degrees of freedom leading to a renormalized interaction potential between all the remaining charge species of the screened DH form,
\begin{equation}
u_{\rm DH}(\Av r,\Av r')=\frac{\rme^{-\kappa \vert\Av r-\Av r'\vert}}{4\pi\varepsilon\varepsilon_0\vert\Av r-\Av r'\vert}.
\label{DH}
\end{equation}
One can thus drop any reference to explicit salt ions and infer thermodynamic
properties of the original system by analyzing it as a system composed
of only ``dressed counterions" (as well as fixed external charges) that
interact via a screened DH pair potential. 
We term this approximation scheme in the SC limit 
as the {\em SC dressed counterion theory}. 
Our SC analysis thus proceeds in the same way as for the 
counterion-only systems \cite{Netz} except that the interactions
between the charges are dressed (Sections \ref{sec:SC-canon} 
and \ref{sec:SC-grandcanon}). 


We note again that on this level the salt ions  are always considered to be weakly coupled to all the other charges as well as between themselves. Any Bjerrum pairing \cite{bjerrum,vanroij} or even electrostatic collapse of the salt or formation of salt-counterion complexes \cite{Fisher} is thus beyond the approach developed here.  
The regime of  parameters  where these effects may become important and
the regime of validity of the SC dressed counterion theory will be
determined later. 

\section{SC dressed counterion theory: Canonical ensemble}
\label{sec:SC-canon}

We first discuss the SC limit within a canonical ensemble for the polyvalent counterions. This means that the 
number of counterions in the system  is fixed. The analysis in this case is very similar to the traditional SC approach \cite{Netz}. In the SC limit, we 
expand the grand-canonical partition function associated with the dressed counterion approximation, Eq. (\ref{hamiltonian-renorm}), 
to the first order in counterion fugacity, $\Lambda_c$, and then perform an inverse Legendre transformation (where the fugacity is mapped to the
number of counterions, $N$, via the relation $\Lambda_c\partial \ln {\mathcal Z}_G/\partial \Lambda_c = N$) in order 
to obtain the canonical SC free energy \cite{Naji-EPJE,Naji,hoda,Netz}. For the system under consideration we find
\begin{equation}
\beta{\cal F}_{N}=\beta W_{00}-N\,\ln\int \rme^{-\beta W_{0c}(\Av r)}\rmd \Av r,
\label{SCfree}
\end{equation}
where the first term is the screened interaction energy of fixed charges (macroions) 
\begin{equation}
W_{00}=\cfrac{1}{2}\int\!\!\!\int\rmd \Av r\rmd \Av r' \rho_0(\Av r)u_\textrm{DH}(\Av r,\Av r')\rho_0(\Av r'),
\label{eq:W_00}
\end{equation}
and the term in the exponent is 
the single-particle interaction energy of the dressed counterions with fixed macroion charges
\begin{equation}
W_{0c}(\Av r)= e_0 q \!\int\rmd \Av r' \rho_0(\Av r')u_\textrm{DH}(\Av r,\Av r').
\label{W0c}
\end{equation}
The SC attraction between like-charged macroions stems from the second term in Eq. (\ref{SCfree}), which contains the counterion-induced effects \cite{Naji,Naji-EPJE,Netz}. 

We now focus on the case of two charged planar surfaces as defined in Section \ref{sec:model}. 
In the canonical ensemble, we stipulate that the number of counterions $N$ between the bounding surfaces remains fixed even as the distance between the surfaces is changed.
The number of counterions, $N$, may be arbitrary and, unlike the counterion-only case \cite{Netz}, its value is not determined by the electroneutrality condition.  
The DH salt takes care in neutralizing all charges as follows from the observation that the dressed interaction potentials are short ranged.

Expressing the interaction energies  with dimensionless quantities, Eq. (\ref{dimless}), and defining 
the rescaled energy $\widetilde W_{00}/\tilde S \equiv (2\varepsilon\varepsilon_0 W_{00})/(\sigma^2\mu_{\mathrm{GC}} S)$  per rescaled area $\tilde S = S/\mu_{\mathrm{GC}}^2$, 
we obtain 
\begin{eqnarray}
\widetilde W_{00}/\tilde S&=&\frac{1}{\tilde \kappa}\,\rme^{-2\tilde\kappa \tilde a},
\label{eq:W_00_2}
\\
\beta W_{0c}(\tilde z)&=&\frac{2}{\tilde\kappa}\,\rme^{-\tilde\kappa \tilde a}\cosh\,\tilde\kappa \tilde z, 
\label{eq:W_0c_2}
\end{eqnarray}
and thus the following  expression for
the SC free energy of the plane-parallel system, 
\begin{equation}
\tilde {\cal F}_{N}/\tilde S=\frac{1}{\tilde \kappa}\,\rme^{-2\tilde\kappa \tilde a}-2\eta\,\ln\,I(\tilde a),
\label{FSC}
\end{equation}
where we have introduced
\begin{equation}
I(\tilde a)=\int_0^{\tilde a}\exp\Bigl(\frac{2}{\tilde\kappa}\,\rme^{-\tilde\kappa \tilde a}\cosh\,\tilde\kappa \tilde z\Bigr)\,\rmd\tilde z.
\label{I}
\end{equation}
The first term in the free energy corresponds to the usual salt-mediated DH repulsive interaction between the surfaces, and the second one is the  contribution of counterions and is thus proportional to their amount in the slit expressed by the dimensionless parameter $\eta$, Eq. (\ref{eta}). The dressed counterion
(number) density profile is obtained in the form of the Boltzmann factor of the single-particle potential 
$W_{0c}(\Av r)$ (i.e., the integrand in Eq. (\ref{I})), and follows as 
\begin{equation}
c(\tilde z)=C\,\exp\Bigl(\frac{2}{\tilde\kappa}\,\rme^{-\tilde\kappa \tilde a}\cosh\,\tilde\kappa \tilde z\Bigr).
\label{c}
\end{equation}
The normalization constant $C$ can be deduced from the known amount of counterions, $\eta$.
The canonical SC theory contains two free physical parameters, namely, the salt screening parameter $\tilde\kappa$ and the parameter 
$\eta$. The model itself additionally contains the coupling parameter, $\Xi$, which is considered  as $\Xi\to\infty$ within the SC theory. We will use MC simulations to compare the results of the SC theory with those obtained numerically at finite values of $\Xi$.

\begin{figure*}[t]
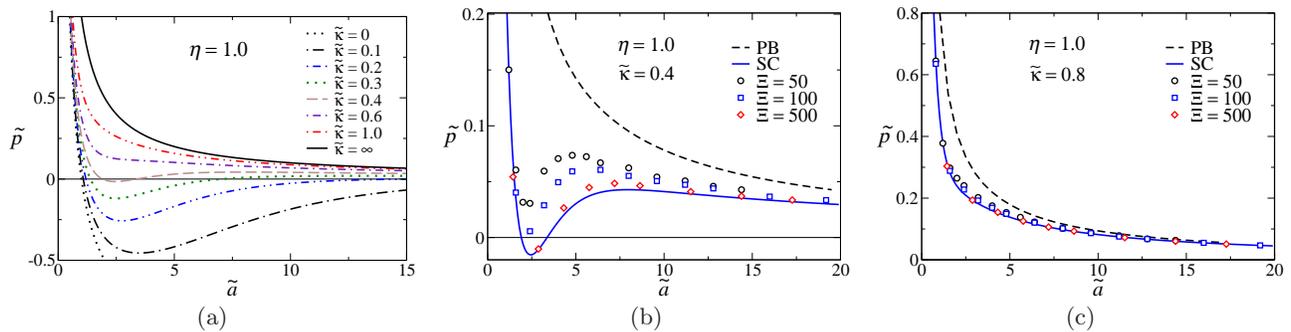
\begin{center}
	\begin{minipage}[b]{0.302\textwidth}\begin{center}
		\includegraphics[width=\textwidth]{picp.eps} (a)
	\end{center}\end{minipage} \hskip0.25cm
	\begin{minipage}[b]{0.302\textwidth}\begin{center}
		\includegraphics[width=\textwidth]{press04-1.eps} (b)
	\end{center}\end{minipage} \hskip0.25cm
	\begin{minipage}[b]{0.302\textwidth}\begin{center}
		\includegraphics[width=\textwidth]{press08-1.eps} (c)
	\end{center}\end{minipage} \hskip0.25cm
	\caption{ (Color online) 
	Rescaled canonical  pressure as a function of the rescaled inter-surface half-distance $\tilde a$. In
	(a) the SC results obtained from Eq. (\ref{p_C}) are shown for $\eta=1$ and different values of the screening parameter $\tilde\kappa$. In (b) and (c) the SC
	result (solid line, Eq. (\ref{p_C}))  and the PB prediction (dashed line, Eq. (\ref{eq:PB_DH_C})) are compared with MC simulation data (symbols) obtained 
	for  different values of the coupling parameter and at fixed $\tilde\kappa$ 
	and $\eta$ as indicated on the graphs.  }
	\label{fig_press1}
\end{center}\end{figure*}

The dimensionless pressure acting on each surface  can be obtained from the free energy via the standard thermodynamic equalities in the form $\tilde p=-\frac{1}{2}\,\partial(\tilde{\cal F}_{N}/\tilde S)/(\partial \tilde a)$, thus yielding 
\begin{equation}
\tilde p=\rme^{-2\tilde\kappa \tilde a}+\eta\,\frac{I'(\tilde a)}{I(\tilde a)},
\label{p_C}
\end{equation}
where the prime denotes the derivative with respect to the argument. The dependence of the SC pressure on the inter-surface separation and the screening length
will be discussed in detail later (Fig.~\ref{fig_press1}a). Here we briefly consider its limiting behavior at small and large separations. 

In the limit of small separations $\kappa a\ll 1$, (i.e., small compared to the screening length $\kappa^{-1}$),  the dimensionless pressure reduces to
\begin{equation}
\tilde p\simeq\frac{\eta}{\tilde a}+(1-2\eta).
\label{psmall}
\end{equation}
The leading order term corresponds to the ideal-gas osmotic pressure of counterions which is repulsive and dominates over other electrostatic contributions.
In the limit of large separations $\kappa a\gg 1$ (i.e., large compared to the screening length $\kappa^{-1}$),  the pressure behaves as
\begin{equation}
\tilde p\simeq\frac{\eta}{\tilde a}-\frac{\eta}{(\tilde\kappa\tilde a)^2},
\label{plarge}
\end{equation}
which indicates that at large separations the counterions also behave as an ideal gas since all their electrostatic interactions are effectively  screened out at several $\kappa^{-1}$ distances. Hence, only the repulsive osmotic contribution remains.

The canonical dressed counterion theory can be used to derive the well-known SC pressure in the no-salt limit $\tilde p_0=1/\tilde a-1$ \cite{Netz}, 
where only (polyvalent) counterions are present without any salt ions. 
This limit is recovered by letting $\kappa\to 0$ and setting $\eta=1$ in order to satisfy the electroneutrality condition. 

\subsection{Regime of validity of the canonical SC dressed counterion theory}

The regime where the SC dressed counterion theory is applicable follows from a simple criterion that was derived and confirmed by MC simulations \cite{hoda,Naji,Netz}. The pivotal statement in deriving this criterion is that the separation between nearest-neighbor counterions must be much larger than the separation between the surfaces, $a_\perp\gg a$. In dimensionless units the separation between neighboring counterions is approximately $\tilde a_\perp\simeq\sqrt{\Xi/\eta}$. The inter-surface separation at which our SC theory should be valid is then
\begin{equation}
\tilde a\ll \sqrt{\Xi/\eta}.
\label{crit_C}
\end{equation}    
This indicates that the regime of validity of the SC theory increases as the coupling parameter becomes larger. Interestingly, the regime of validity of the theory increases 
also as  $\eta$ becomes smaller (at fixed coupling). In particular, for $\eta<1$ (i.e., when
the amount of the bare charge due counterions is less than the bare fixed charge on the macroions) the SC dressed counterion theory is expected
to hold in a wider range of separations as compared with the original no-salt SC theory \cite{Netz}. 

\begin{table}
	\begin{center}
\begin{tabular}{c|lllll}
$\Xi$	&	$\sigma$	&	$q~\quad$	&	$n_0^{min}$~[M]&	$n_0$~[M]~	&	$c_0$~[M]\\
	&	[$e_0$/nm$^2$]~	&			&	($\tilde\kappa=1/q)$&	($\tilde\kappa=1$)&($\tilde\chi=1$)\\
\hline
1	&	0.3	&	1	&0.2	&	0.2	&	0.2\\
\vspace{-1.5ex}\\
50	&	0.24&	4	&0.1	&	1.7	&	0.1\\
		&	0.13	&	5	&0.03	&	0.8	&	0.03\\
\vspace{-1.5ex}\\
100	&	0.15	&	6	&0.04	&	1.5	&	0.04\\
	&	0.1	&	7	&0.02	&	0.8	&	0.02\\

\vspace{-1.5ex}\\
500 &	0.06	&	7	&0.02	&	1.0	&	0.02\\
(ethanol)	&		&	&		&		&	\\
\end{tabular}
\caption{Illustrative examples of possible physical parameters corresponding to a given set of values for the electrostatic coupling parameter $\Xi$
 (in water at room temperature with $\ell_{\mathrm{B}}\approx 0.7$ nm). $n_0^{min}$ shows the minimal salt concentration in order to satisfy the criterion (\ref{crit_DH}).
The salt concentration $n_0$ corresponds to the inverse dimensionless screening length $\tilde\kappa=1$ and the bulk polyvalent counterion concentration $c_0$ corresponds to $\tilde\chi=1$. For $\Xi=500$ water as the dielectric medium ($\varepsilon=80$) is replaced by ethanol ($\varepsilon=30$).
\label{table}
} 
\end{center}
\end{table}

On the other hand, at very large inter-surface separations, where most of the electrostatics is screened out, the interaction between counterions becomes negligible and the SC theory of the dressed counterions, Eq. (\ref{FSC}), retains its validity again. We therefore expect that the SC for dressed counterions ceases to be valid only at intermediate separations.

The validity of the DH-type linearization  that we have used to derive the dressed interaction potentials is also limited by stipulating that the dimensionless DH potential itself is always small enough. Specifically this sets the value of the dimensionless DH potential at the two boundaries to be well below unity, which translates to the condition $e_0 \kappa\gg 2\pi\ell_{\mathrm{B}}\sigma$, or 
\begin{equation}
\tilde\kappa\gg 1/q,
\label{crit_DH}
\end{equation}  
where $q$ is the counterions valency. 
Table \ref{table} shows illustrative examples of possible physical parameters where the above criterion is fulfilled. 

\begin{figure}[t]
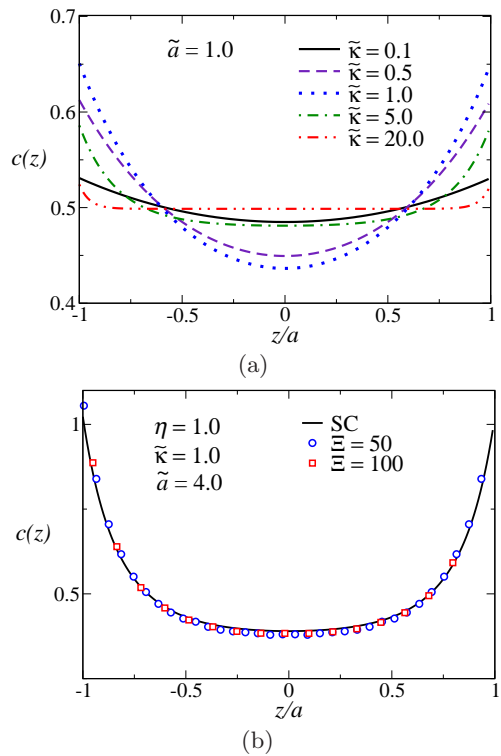
\begin{center}
\begin{minipage}[b]{0.36\textwidth}\begin{center}
\includegraphics[width=\textwidth]{picrho1.eps} (a)
\end{center}\end{minipage} \hskip0.25cm
\vskip0.1cm
\begin{minipage}[b]{0.36\textwidth}\begin{center}
\includegraphics[width=\textwidth]{dens-1-1.eps} (b)
\end{center}\end{minipage}
\caption{(Color online) a) Normalized SC dressed counterion density profile between two charged planar surfaces within the 
canonical description, Eq. (\ref{c}), for half-distance $\tilde a=1$ and various screening parameters $\tilde\kappa$. The normalized SC density is independent of the parameter $\eta$. b) The theoretical SC density 
profile (solid line) is compared with MC simulation data (symbols)
 for $\tilde \kappa=1$, $\eta=1$, $\tilde a=4$ and two different coupling parameters $\Xi$ as used in the simulations. }
\label{dens_results}
\end{center}\end{figure}

\begin{figure*}[t]
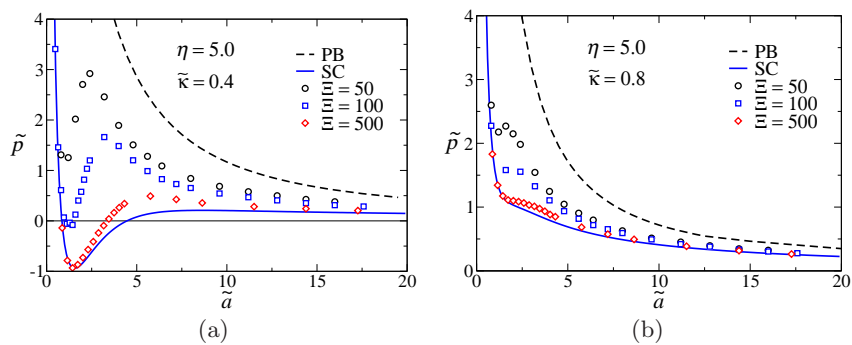
\begin{center}
	\begin{minipage}[b]{0.302\textwidth}\begin{center}
		\includegraphics[width=\textwidth]{press04-5.eps} (a)
	\end{center}\end{minipage} \hskip0.25cm
	\begin{minipage}[b]{0.302\textwidth}\begin{center}
		\includegraphics[width=\textwidth]{press08-5.eps} (b)
	\end{center}\end{minipage} \hskip0.25cm
	\caption{(Color online) Same as Fig.~\ref{fig_press1}b and c but for $\eta=5.0$.} 
	\label{fig_press2}
\end{center}\end{figure*}

\subsection{MC simulations: Canonical ensemble}

In order to assess the validity of the  SC dressed counterion theory, we performed MC simulations and compared them with the analytical results. Simulations cover
a wider range of coupling parameters and can thus fill the gap between the WC and the SC coupling limits investigated analytically in the
preceding Sections. The simulation details are described in Appendix \ref{app:sim}. 

First we compare the counterion density profiles obtained from the simulations with the analytical result, Eq. (\ref{c}).
As seen in Fig.~\ref{dens_results}, the counterion density profile is a convex function of the inter-surface position so that the counterions are preferentially gathered near both charged surfaces due to electrostatic attraction. For large screening or large separations, $\kappa a\gg 1$, the typical counterion density decay length from each surface is $\kappa^{-1}$. In contrast, at low screening or small separations, $\kappa a\ll 1$, the typical decay length is $\sqrt{\kappa^{-1}\mu_{\mathrm{GC}}}$. 
Note that the density profile has non-monotonic variations as the screening is increased. For vanishing $\kappa$ \cite{Netz} as well as for very large $\kappa$ the SC dressed counterion density profile becomes homogeneous, $\rho(z)=\textrm{const}$.
The agreement with MC data is excellent already at $\Xi=50$ as shown in Fig.~\ref{dens_results}. 

Next we consider the inter-surface pressure. The analytical pressure-distance plots from Eq. (\ref{p_C}) are shown in  Fig.~\ref{fig_press1}a for a few different screening parameters. 
At small separations the pressure is always repulsive due to the large osmotic component of counterions stacked in the small region between the surfaces and is  independent of the screening, Eq. (\ref{psmall}). By increasing the inter-surface separation, the osmotic pressure as well as the electrostatic counterion-induced attraction (embedded in the second term of
Eq. (\ref{SCfree}) or (\ref{p_C})) are reduced. The repulsive osmotic pressure decays approximately as $1/a$ for small distances, whereas attractive electrostatic contributions decay with the  length scale  $\kappa^{-1}$. Thus, when $\kappa$ is small enough, the attractive electrostatic component may remain large enough at intermediate distances to overcome osmotic repulsion and hence a net attraction can appear.

As clearly seen form Fig.~\ref{fig_press1}a, the larger the screening parameter $\kappa$ the smaller will be the attraction at intermediate separations. For large enough $\kappa$ the attraction disappears altogether and the pressure reverts to its limiting form, Eq. (\ref{plarge}). For large inter-surface separations the pressure thus always ends in a repulsive regime due to slow ($1/a$) decay 
of the osmotic contributions as compared to rapid ($\rme^{-\kappa a}$) decay of the electrostatic contributions. 

Comparison with MC results in Figs. \ref{fig_press1}b and c shows expectedly that the agreement becomes better as the coupling parameter $\Xi$ becomes larger. The agreement is also better for a smaller fraction of counterions, $\eta$, in the slit (cf. Figs.~\ref{fig_press1} and \ref{fig_press2}), which also follows from Eq. (\ref{crit_C}). At large separations or strong screening the electrostatics is screened out and the SC dressed counterion approximation becomes valid again. 

The validity of the SC approximation at large separation or strong screening is a consequence of the dressed counterion theory and is not obtained in the standard SC theory with counterions only, which remains valid only in the regime of small separations \cite{Naji,Netz,hoda}. In the dressed counterion approach at high screening or large separations the counterions behave as   ``ideal" 
 particles   and do not interact anymore with fixed charges in the system leading to a repulsive  ideal-gas pressure regime with the $1/a$ signature dependence on the inter-surface separation. 
This dependence is not obtained in the counterion-only SC theory where the interactions remain attractive and long range at large separations. 
We may thus conclude that the SC dressed counterion theory captures the physics both at large and small separations and requires improvements only at intermediate distances.

\begin{figure}[t]\begin{center}
\begin{minipage}[b]{0.38\textwidth}\begin{center}
\includegraphics[width=\textwidth]{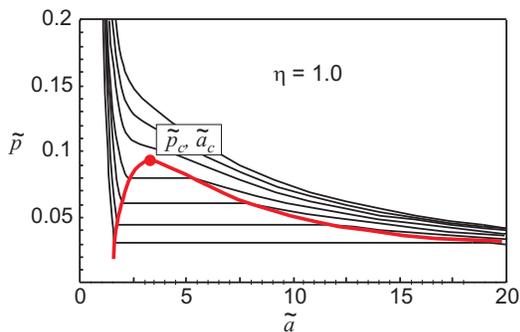}
\end{center}\end{minipage} \hskip0.25cm
\caption{(Color online) Van der Waals type iso-inverse screening length curves are shown
along with the corresponding Maxwell construction for  the interaction pressure between charged plates within the canonical description for dressed counterions. Here $\eta = 1$ and the dimensionless inverse screening length varies in the range from $\tilde\kappa = 0.4$ to $\tilde\kappa = 0.56$ in intervals of 0.027
 (from bottom to top). Red line represents the critical curve. The critical point is located at approximately $\tilde p_c = 0.092, \tilde a_c = 3.07$ and $\tilde\kappa_c = 0.546.$}
\label{figSC}
\end{center}\end{figure}

The interaction pressure in the canonical system thus always possesses repulsive branches at small and large separations and can show non-monotonic behavior in between
as seen in the figures.  In fact, for certain values of the parameters the interaction pressure 
shows a van der Waals-like loop which could suggest a coexistence regime between two different   ``phases". The difference between the standard van der Waals isotherms and the case encountered here is that the curves are not isotherms but are rather iso-inverse screening length or  iso-ionic strength curves. From a thermodynamic perspective, this would indicate a dense phase identified with a small inter-surface separation at equilibrium with an expanded phase with a large separation.  
Such van der Waals-type loops in the interaction pressure between interacting charged surfaces have been seen in other contexts before \cite{Orr}. 
The phase coexistence can be demonstrated by means of a Maxwell construction analysis as shown in Fig.~\ref{figSC}. 
In an experiment that can only probe stable equilibrium states in the system, as is the case in osmotic stress experiments and in general also in the surface force experiments \cite{Israelachvili}, one would measure exactly these type of interaction pressure {\em versus} separation curves that are in agreement with the appropriate Maxwell construction.
The binodal or the coexistence curve, that delimits the region in the pressure-separation dependence where Maxwell construction is feasible, ends at a critical point corresponding to the largest amount of salt in the system that still leads to a non-monotonic dependence of the pressure on the inter-surface separation. Above this limiting salt concentration the interaction pressure is purely repulsive. For the case with $\eta=1$ we numerically obtain the critical point as $\tilde p_c=0.092$, $\tilde a_c=3.07$ and $\tilde\kappa_c=0.546$ in rescaled units.

It is interesting to note that this type of interaction pressure equilibria corresponding to abrupt transitions from one equilibrium separation to a different one are indeed {\em observed} between strongly charged macromolecules in the 
presence of polyvalent counterions and monovalent salt. Most notably osmotic stress experiments on DNA in the 
presence of trivalent CoHex counterions and 0.25M NaCl salt show that for intermediate ${\rm Co}^{3+}$ 
concentrations (8 mM and 12 mM) there are abrupt transitions from one repulsive osmotic pressure branch to 
another one \cite{rau-1}. Similar features are discerned even for a divalent counterion  ${\rm Mn}^{2+}$ at various concentrations or temperatures \cite{rau-2}.

\section{SC dressed counterion theory: Grand-canonical ensemble}
\label{sec:SC-grandcanon}

So far we have considered only the canonical ensemble where the system contains a fixed amount of dressed (polyvalent) counterions. In this Section we focus on the grand-canonical ensemble for the SC dressed counterion theory. In the grand-canonical ensemble, the system is in equilibrium with an external reservoir of dressed counterions with the bulk concentration $c_0$. Depending on a chemical potential (or the fugacity $\Lambda_c$), the system can adjust the number of counterions by exchanging them with the reservoir. The number of counterions, $N$, in the  inter-surface region thus varies with the distance between the surfaces.

Here we will derive the grand-canonical free energy by using the canonical free energy, Eq. (\ref{FSC}), discussed in the previous Section.
The  grand canonical partition function is defined as
\begin{equation}
\rme^{-\beta \Phi} \equiv {\mathcal Z}_G =\sum_{N=0}^{\infty}\frac{\Lambda_c^N}{N!}{\cal Z}_N, 
\end{equation}
where  ${\cal Z}_N=\rme^{-\beta{\cal F}_N}$ is the canonical partition function. 
The grand-canonical free energy is then obtained as 
\begin{equation}
\beta\Phi=\beta W_{00} - \Lambda_c I(a),
\end{equation}
where $W_{00}$ is the screened interaction energy of the charged surfaces (in the absence of counterions), 
Eqs. (\ref{eq:W_00}) and (\ref{eq:W_00_2}), and $I(a)$ is given by Eq. (\ref{I}). We can evaluate the average number of counterions in the slit as
\begin{equation}
\overline N=-\Lambda_c\left(\frac{\partial \beta\Phi}{\partial \Lambda_c}\right)_{\beta, a}=\Lambda_c I(a), 
\label{eqN1}
\end{equation}
and the pressure is given in terms of this latter quantity as 
\begin{eqnarray}
p_{\rm sub}&=&-\left(\frac{\partial\Phi}{\partial V}\right)_{\beta, \Lambda_c}\\
	&=&-\frac{1}{2S}\left[\frac{\partial W_{00}}{\partial a}-\overline N k_{\mathrm{B}}T \frac{\partial}{\partial a}\ln\,I(a)\right].
\label{eq_p_grand}	
\end{eqnarray}
In order to satisfy the chemical potential equivalence with the reservoir, counterions must be distributed in the inter-surface region according to the Boltzmann weight,
\begin{equation}
c(\Av r)=c_0\,\rme^{-\beta W_{0c}(\Av r)}.
\end{equation}
Here $W_{0c}(\Av r)$ is the single-particle surface-counterion interaction energy, Eqs. (\ref{W0c}) and
(\ref{eq:W_0c_2}). The rescaled dressed counterion density in the inter-surface region is then obtained as 
\begin{equation}
c(\tilde  z)=c_0\exp\Bigl(\frac{2}{\tilde\kappa}\,\rme^{-\tilde\kappa \tilde a}\cosh\,\tilde\kappa \tilde z\Bigr).
\label{c_GC}
\end{equation}
Again this single-particle form of the density distribution is expected in the SC limit as counterion-counterion interaction effects are negligible on the leading order \cite{Netz,hoda,Naji}. The density profile in the grand canonical (Eq. (\ref{c_GC})) and canonical ensembles (Eq. (\ref{c})) have similar forms, except that the normalization constants are different. In the present case, 
$c_0$ is the bulk concentration of dressed counterions and is therefore a free parameter. 
Integrating the counterion concentration in the inter-surface region, we get the average number of counterions as
\begin{equation}
\overline N=\int c(\Av r)\, \rmd V= S\mu_{\mathrm{GC}}\, c_0 I(a).
\end{equation}

The expression (\ref{eq_p_grand}) for the pressure includes only the contribution due to the free energy of the ions in the system. 
In order to evaluate the net pressure acting on each surface, one must also account for the pressure exerted by the counterions in the bulk (reservoir), i.e.  
\begin{equation}
p_{{\rm ci}0}=c_0 k_BT.
\end{equation}
The net pressure,  $p=p_{\rm sub}-p_{{\rm ci}0}$,  is then obtained as 
\begin{equation}
\tilde p=\rme^{-2\tilde\kappa\tilde a}+\frac 18 \tilde\chi^2 [I'(\tilde a)-2]
\label{p_GC}
\end{equation}
in rescaled units, where $\tilde \chi^2=8\pi q^2 \ell_{\mathrm{B}} c_0\,\mu_{\mathrm{GC}}^2$. 
The grand-canonical  theory therefore contains two free physical parameters, i.e., $\tilde\kappa$ and $\tilde \chi$, which 
are given by the bulk salt and polyvalent counterions concentrations, respectively, as well as the rescaled half-distance, $\tilde a$.

We can re-express the above pressure explicitly as
\begin{equation}
\tilde p=\rme^{-2\tilde\kappa\tilde a}+\frac 14 \tilde \chi^2 \Bigl[\rme^{-\beta W_{0c}(\tilde a)}-1-\int_{-\tilde a}^{\tilde a}\rme^{-\tilde\kappa(\tilde z+ \tilde a)}\rme^{-\beta W_{0c}(\tilde z)}\rmd \tilde z\Bigr],
\label{p_GC1}
\end{equation}
which allows for a simple physical interpretation. The first term is the usual DH repulsion between the plates, exactly the same as in the canonical case, Eq. (\ref{p_C}). The remaining part, proportional to $\tilde \chi^2$, is the dressed counterions contribution, i.e. 
\begin{equation}
\tilde p_c=\frac 14 \tilde \chi^2 \Bigl[\rme^{-\beta W_{0c}(\tilde a)}-1-\int_{-\tilde a}^{\tilde a}\rme^{-\tilde\kappa(\tilde z+ \tilde a)}\rme^{-\beta W_{0c}(\tilde z)}\rmd \tilde z\Bigr].
\label{p_GC1_c}
\end{equation}
 The first two terms in the brackets correspond to the difference between counterion osmotic pressures inside the system and in the bulk, which is hence repulsive. The last term in the brackets contains the attractive electrostatic interaction mediated by dressed counterions between the two surfaces. 
 
 The pressure at small separations, $\kappa a\ll 1$, varies  
linearly with the distance as
\begin{equation}
\tilde p\simeq 1+\frac 14 \tilde \chi^2(\rme^{2/\tilde\kappa}-1)-(2\tilde\kappa+\tilde \chi^2\rme^{2/\tilde\kappa})\,\tilde a.
\label{p_GC2}
\end{equation}
Note that, in contrast to the canonical case, the grand-canonical pressure reaches a finite value as the separation tends zero, since in this limit all counterions are expelled from the inter-surface region. 
For large separations, $\kappa a\gg 1$, the pressure approaches zero asymptotically from the negative side
\begin{equation}
\tilde p\sim -\frac 12\frac{\tilde \chi^2}{\tilde \kappa}\,\tilde a\,\rme^{-2\tilde \kappa \tilde a}.
\label{p_GC3}
\end{equation}
The above limiting  form involves contributions from electrostatic as well as entropic effects that both tend to zero at large separations.

In contrast to the canonical case, discussed in the previous Section, the grand-canonical case has no counterion-only analog that could be obtained simply by taking the limit $\kappa\to 0$. The reason is that the amount of counterions between the surfaces in the grand-canonical ensemble is controlled by the inter-surface potential and hence by the amount of all ions between the surfaces. 
Since in the limit $\kappa\to 0$, only counterions are present, 
the electroneutrality which implies a fixed amount of counterions 
cannot be generally achieved. The conclusion is that the grand-canonical ensemble cannot be used to study the low-salt (or zero salt) limit, where a canonical description is more
appropriate.

\subsection{Regime of validity of the grand-canonical SC dressed counterion theory}

The amount of counterions between the surfaces is fixed in the canonical case and can be expressed in terms of the parameter $\eta$, Eq. (\ref{eta}). In the grand-canonical case counterions can be exchanged with an external reservoir. Defining the grand-canonical fractional amount of counterions, $\eta_{\rm GC}$, in the slit in accordance with the definition (\ref{eta}), we find that 
\begin{equation}
\eta_\textrm{GC}(\tilde a)=\frac 18 \tilde \chi^2 I(\tilde a).
\label{etaGC}
\end{equation}

In order to obtain a validity criterion for the grand-canonical SC theory, we follow the same general argument as used in the canonical case, i.e., the SC
description is expected to hold when $\tilde a\ll \tilde a_\bot$, and hence $\tilde a\ll\sqrt{\Xi/\eta_\textrm{GC}}$. As discussed before, $\eta_\textrm{GC}$ now depends on the separation $a$. Using Eq.  (\ref{etaGC}), we obtain the following criteria
\begin{equation}
	\begin{array}{ll}
	\tilde a\ll\cfrac{\Xi^{1/3}}{\tilde \chi^{2/3}}\,\rme^{-2/3\tilde \kappa} & \quad\textrm{ for }\kappa a\ll 1,\\
	&\\
	\tilde a\ll\cfrac{\Xi^{1/3}}{\tilde \chi^{2/3}} & \quad\textrm{ for }\kappa a\gg 1,
	\end{array}
	\label{crit_GC}
\end{equation}
which clearly merge into a single criterion if $\tilde\kappa\gg 1$. These criteria determine the range of parameters where the grand-canonical SC dressed counterion theory 
may be applied at {\em finite} couplings, e.g., when the theory is compared with numerical simulations. Note that the range of the validity of the grand-canonical theory is more limited
than that of the canonical theory as the dependence on the coupling parameter, $\Xi$, turns out to scale as $\Xi^{1/3}$ in contrast to the usual $\Xi^{1/2}$
behavior found in the canonical case or within the standard no-salt SC theory \cite{Netz,Naji,hoda}. But  the regime of validity of the theory increases  
as the bulk  concentration of counterions is lowered, which is similar to the trend found within the canonical theory.

\subsection{MC simulations: Grand-canonical ensemble}

\begin{figure}[t]\begin{center}
	\begin{minipage}[b]{0.39\textwidth}\begin{center}
		\includegraphics[width=\textwidth]{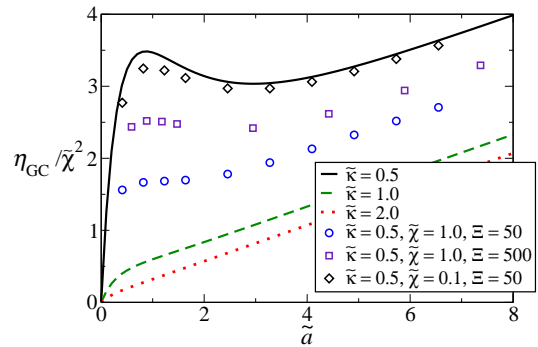}
	\end{center}\end{minipage} \hskip0.25cm

\caption{ (Color online) Charging curves of the grand-canonical system: the rescaled 
amount of counterions in the slit, $\eta_\textrm{GC}/\tilde \chi^2$, is plotted 
as a function of the rescaled half-distance $\tilde a$ for various screening parameters $\tilde\kappa$. 
The theoretical SC values of $\eta_\textrm{GC}/\tilde \chi^2$ (lines)  are independent of $\tilde \chi$. Symbols correspond to MC results for $\tilde\kappa=0.5$ which indicate better agreement with 
the theoretical prediction for larger $\Xi$ and smaller $\tilde \chi$.}
	\label{fig_eta}
\end{center}\end{figure}

An interesting quantity that can be investigated in the grand-canonical case is the amount of counterions, $\eta_\textrm{GC}$, between the surfaces as the inter-surface separation, $2a$, is varied. Since $\eta_\textrm{GC}$ depends linearly on the amount of counterions in the bulk, the quantity $\eta_\textrm{GC}/\tilde \chi^2$ depends only on the screening parameter $\kappa$ in addition to the 
inter-surface separation, $2a$.
In Fig.~\ref{fig_eta}, we show the ``charging" curve, $\eta_\textrm{GC}/\tilde \chi^2 = \eta_\textrm{GC}/\tilde \chi^2(\tilde a)$, which reflects some of the properties of the counterion density profile. The amount of counterions starts to increase linearly at small separations, $\kappa a\ll 1$. This is due to the fact that at small separations counterion density profile is almost constant in the inter-surface region  and thus the number of counterions is proportional to the inter-surface distance.
On the other hand, at very large separations, $\kappa a\gg 1$, the counterion density profile is again very flat in the middle of the inter-surface region, which again results in a linear dependence of the amount of counterions on the  inter-surface distance as shown in the figure. 

It is interesting to note that for screening parameters smaller than $\tilde\kappa<0.575$, the charging process is non-monotonic in an intermediate
range of  separations. This is a consequence of the interplay between the inter-surface separation and the resulting ``dressed" electrostatic potential. In the non-monotonic part of $\eta_\textrm{GC}$, the increase in the separation results in a larger drop of the resulting DH potential between both surfaces. The corresponding potential drop then results in a larger decrease in the
number of counterions than would follow from the increase of the separation itself and hence counterions are expelled from the inter-surface space as a result of the increase in the inter-surface
distance! 

As seen in Fig.~\ref{fig_eta}, the theoretically predicted value of $\eta_\textrm{GC}$ gives the upper bound limit and the MC results at finite coupling parameter are always smaller. 
The agreement between the theory and simulations is better for larger coupling parameter $\Xi$. In fact as $\Xi$ is decreased the counterion-counterion repulsions 
become more dominant and the amount of counterions in the slit is decreased.
The agreement  is also better for smaller concentration of counterions (smaller $\tilde \chi$), which  implies larger nearest-neighbor distances between the counterions and thus smaller
counterion-counterion repulsions. These observations agree qualitatively with the criteria  (\ref{crit_GC}).

\begin{figure*}[t]
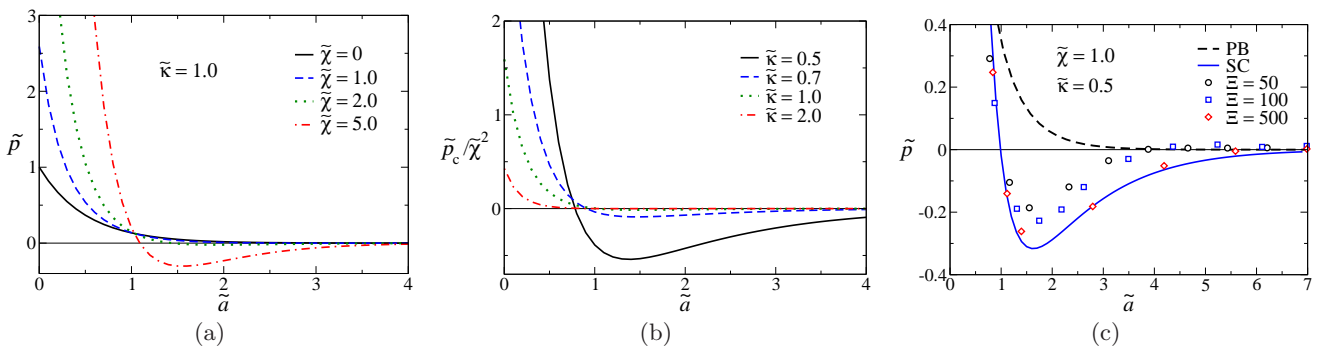
\begin{center}
	\begin{minipage}[b]{0.302\textwidth}\begin{center}
		\includegraphics[width=\textwidth]{press_GC1.eps} (a)
	\end{center}\end{minipage} \hskip0.25cm
	\begin{minipage}[b]{0.32\textwidth}\begin{center}
		\includegraphics[width=\textwidth]{press_GCci.eps} (b)
	\end{center}\end{minipage} \hskip0.25cm
	\begin{minipage}[b]{0.307\textwidth}\begin{center}
		\includegraphics[width=\textwidth]{press_grandMC2.eps} (c)
	\end{center}\end{minipage} \hskip0.25cm
	\caption{ (Color online) 
	Rescaled grand-canonical  pressure as a function of the rescaled inter-surface half-distance $\tilde a$. In
	(a) the SC results obtained from Eq. (\ref{p_GC1}) are shown 
	for $\tilde\kappa=1$ and different values for the parameter  $\tilde  \chi$, which is related to the bulk concentration of dressed counterions as defined in the text.
	In (b) the rescaled counterion contribution, $\tilde p_c/\tilde \chi^2$ (Eq. (\ref{p_GC1_c})), to the inter-surface interaction pressure is shown  for different values of the screening parameter $\tilde\kappa$ as indicated on the graph.
	In (c) the theoretical SC (solid line, Eq. (\ref{p_GC1}))
	and PB (dashed line, Eq. (\ref{eq:P_PB_GC})) results are compared with MC results (symbols)  
	obtained for different values of the coupling parameter and at fixed $\tilde \chi=1$ and $\tilde\kappa=0.5$. }
	\label{fig_pressGC}
\end{center}\end{figure*}

There are some qualitative differences between the SC dressed counterion pressure in the two ensembles. Most notably, 
 the grand-canonical pressure (shown in Fig.~\ref{fig_pressGC}a) does not diverge and has a finite limiting value at small separations, Eq. (\ref{p_GC2}), 
while the canonical pressure diverges due to the diverging contribution of the confinement entropy of counterions. 
 Recall that the total pressure in the grand-canonical case, Eq. (\ref{p_GC1}), is composed of a contribution due to the surface-surface repulsion, which is proportional to $\exp(-2\kappa a)$, and a contribution due to dressed counterions. As seen from Eq. (\ref{p_GC1_c}), the latter contribution, $\tilde p_c$, is proportional to the amount of counterions in the bulk, $\tilde \chi^2$. Hence, $\tilde p_c/\tilde  \chi^2$ depends only on the inverse screening length $\kappa$ and the separation $a$. The behavior of this contribution is shown  in Fig.~\ref{fig_pressGC}b which shows a minimum at rather small separations
whose depth decreases rapidly with increasing the screening parameter $\kappa$. Whether the total pressure exhibits an attraction regime (with negative total pressure) at a given screening parameter $\kappa$ depends 
on the amount of counterions in the system through  $\tilde \chi^2$. 

In Fig.~\ref{fig_pressGC}c, we compare the SC pressure with the  MC simulations in the grand-canonical case. 
There is a good agreement at small separations in accordance with the criteria  (\ref{crit_GC}). 
At large separations better agreement may be obtained by taking a larger value of the coupling parameter.
In the grand-canonical case,  the interaction osmotic  pressure decays exponentially, Eq. (\ref{p_GC3}),  and therefore electrostatic effects, which also decay exponentially, remain important even at large separations. 

As seen the simulation results are again bracketed by the two limiting analytical theories of weak and strong coupling within the dressed counterion scheme
and thus agree with the general feature obtained before \cite{Naji,Netz,hoda} that the WC and SC limits in fact establish the upper and lower bounds for the interaction pressure
between charged surfaces.

\section{Discussion}

\subsection{Comparison between canonical and grand-canonical ensembles}   

In the preceding Sections, we analyzed the behavior of a system of two charged surfaces in the presence of (polyvalent) counterions and simple monovalent salt ions. We proposed two different ensembles for the polyvalent counterions, i.e., the canonical ensemble, where the number of counterions in the system is fixed and determined by the parameter $\eta$, and the grand-canonical ensemble, where the
number of counterions is regulated by the inter-surface separation and via a chemical equilibrium with the bulk reservoir where the counterion concentration is determined by the parameter $\chi$. 
The grand-canonical ensemble does not only quantitatively differ from the canonical one but also exhibits some very important qualitative distinctions that we shall
discuss further in this Section. 

\begin{figure}[t]\begin{center}
	\begin{minipage}[b]{0.36\textwidth}\begin{center}
		\includegraphics[width=\textwidth]{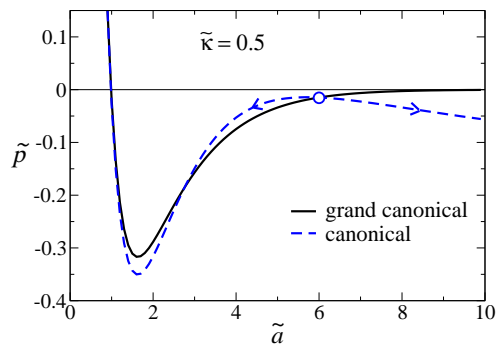}
	\end{center}\end{minipage} \hskip0.25cm
	\caption{(Color online) Rescaled pressure of the grand-canonical system with $\tilde \chi=1$ when equilibrated initially at the inter-surface  half-distance $\tilde a_0$=6 and fixed $\tilde \kappa = 0.5$. For slow changes in the distance between the surfaces the pressure follows the 
	grand-canonical curve (solid black line). For rapid changes in the separation the pressure is expected to follow the canonical curve (dashed blue line).}
	\label{fig_GCvsC}
\end{center}\end{figure}


As noted before the pressure in the canonical case diverges due to large osmotic contribution from counterions in the slit at small inter-surface separations. In contrast, the grand-canonical pressure approaches a finite value as counterions escape from the slit at small separations.

Another distinctive feature is that there is a prominent hump in the canonical case which makes it possible to use a Maxwell construction analysis suggesting a phase 
coexistence behavior. We find no such behavior 
in the grand-canonical case.

The canonical description of our model system may be relevant for two large thin membranes permeable to simple salt (small) ions and impermeable to (large) counterions. It may be assumed 
that both membranes are sealed together at the edges so that the counterions cannot escape from the inter-surface region. 
On the contrary, in the grand-canonical ensemble the system is open and equilibrates with the bulk reservoir. 
In practice one must of course pay attention to the relaxation time scales since polyvalent counterions, being usually bulkier, may typically require 
more time to equilibrate after every change of separation than the small salt ions.

Two equilibration time scales thus unavoidably come into play. At every change of separation counterions first redistribute in the slit, which happens rather quick. Additionally, they also equilibrate with the counterions in the reservoir, but this reservoir equilibration time is usually much larger than the first one, especially if the plates are very large. For time scales smaller than the reservoir equilibration time of counterions to enter the slit, the  system would behave as predicted by the canonical variant of the theory. 

Figure \ref{fig_GCvsC} shows an example of the pressure-distance behavior for a grand-canonical system that was equilibrated  with the bulk  at inter-surface half-distance 
$\tilde a_0=6$. The amount of counterions in the slit is therefore $\eta_{\textrm{GC}}(\tilde a_0)$. Thus any change of separation, after the system is left 
for a sufficiently  long time to relax, results in the grand-canonical pressure curve as shown by the solid line in the figure (Eq. (\ref{p_GC})). 
On the contrary, if the separation changes (starting at the initial value $a_0$) are fast enough, i.e., for time scales smaller than the counterion equilibration time with the reservoir, 
the amount of counterions $\eta$ would remain unchanged as $\eta=\eta_{\textrm{GC}}(\tilde a_0)$. 
Hence the system would behave as a canonical system.
Using the canonical expression (\ref{p_C}) for fixed $\eta$, we can calculate the pressure as $\tilde p(\tilde a)-\frac 14 \tilde\chi^2$ which is
 shown by the dashed line in Fig.~\ref{fig_GCvsC} (note that only the slit region behaves canonically and we still need
to subtract the osmotic bulk contribution of counterions $-\frac 14 \tilde\chi^2$; this term is not present in  Eq. (\ref{p_C}) where no counterions are considered in the bulk).  
When the separation  $a$ is increased  in the canonical approach  from the equilibrium point $a_0$,   the osmotic counterion contribution decreases, 
whereas the outer bulk contribution $-\frac 14 \tilde\chi^2$ remains unchanged. Therefore, the canonical pressure goes to negative values due to the pressure of the counterions from the outside and eventually limits to the bulk counterion pressure $-\frac 14 \tilde\chi^2$. The  deficit  of counterions in the slit will be compensated by counterions from the bulk only after a long enough time, after which the pressure curve would coincide with the grand-canonical result.


\subsection{Finite counterion size}

In order to bring up the principal features of our approach, 
we have assumed here that the counterions are point-like. In general, in the SC limit, counterion-counterion interactions do not enter 
the leading-order SC theory and thus excluded-volume interactions do not affect the leading-order results (reflecting the fact that counterions
are highly isolated in large correlation holes at high couplings) \cite{Netz,Naji,hoda}. However, counterion size effects may become
important when additional salt is present in the system. 

In practice all ions have finite radii, which, for monovalent salt ion, are typically around $0.1$~nm, whereas for polyvalent ions, the ion size (radius) $R_c$ may be as large as  $0.3-1.0$~nm or even larger depending on the ion type. Since the (polyvalent) counterions can thus be much larger we will give some remarks on the possible influence of the counterion size on our results \cite{note-multipole}. The first effect is the hard-core repulsion of counterions from the surfaces. This effect is taken into account {\em exactly} within the leading-order
SC theory \cite{Naji,Naji-EPJE}  via the closest approach distance between the surfaces which enters as an excluded volume in the integrals (\ref{I}) and (\ref{p_GC1}). This would shift the interaction pressure curves by $2R_c$ to the right. On top of that depletion effects due to the exclusion of counterions would be visible in the grand-canonical case for inter-surface separations below $2R_c$ since in that case the counterions cannot enter the inter-surface region \cite{Matej-cyl}.

Note also the electrostatic correction introduced by the finite counterion size may be incorporated within the DH scheme employed here simply via  the 
well-known charge renormalization factor  $\exp(\kappa a)/(1+\kappa a)$. For counterion radius in the range from $0.2~$nm to $0.4~$nm and $\kappa$ from $0.5~$nm$^{-1}$ to $2~$nm$^{-1}$, this would give $\kappa a$ in the range  $0.1$ to $1$, and thus the renormalized counterion charge is varied only by a factor between $1.1$ and $1.4$. 
This charge valency renormalization therefore leads to 
only small changes in the corresponding value of the
coupling parameter $\Xi$ (which is proportional to $q^3$) in the range $1.3$ to $3$.
Another related effect is the failure of the linear DH screening in the vicinity of polyvalent counterions due to high values of the electrostatic potential \cite{Kjellander}. As a result, the electrostatic potential in the vicinity of the counterions decays faster than predicted by the linear screening theory. At larger distances from the counterion the electrostatic potential is again DH-like but with a different effective charge value that can be captured  by an effective valency, $q_\textrm{eff}$. Typically if the counterion valency and radius $R_c$ 
are such  that  $q\ell_{\mathrm{B}}/R_c\lesssim 10$, the effective charge is of the order of the bare charge. But above this threshold, the effective charge gradually becomes smaller than the actual charge and eventually saturates at some fixed 
value $q^\textrm{sat}_\textrm{eff}$ \cite{alex,trizac}. For typical polyvalent counterion size $R_c\sim 0.6$~nm, the effective charge renormalization becomes important for $q\gtrsim 9$. 

The excluded-volume repulsion between counterions results also from their finite size and leads to structuring of the Coulomb fluid at high counterion densities. These effects are however well known and are not distinctive for the system at hand. We thus do not venture into the details of the hard-core structuring between hard walls since it has been amply discussed elsewhere \cite{Israelachvili}.

There are several other effects that come into play because of the ion-ion interactions and correlations such as the Bjerrum pair formation which is another manifestation of the non-linear screening effects \cite{bjerrum, vanroij}. Such effects were studied separately from corrections to traditional PB and DH approaches. They are expected to be important again for $q\ell_{\mathrm{B}}/R_c > 10$, which falls well outside the regime of validity of the dressed counterion theory that we consider here. Thus we expect that these finite size effects lead to no new physics and no new qualitative features emerge in the behavior of the system in the regime relevant to our study. The details of their possible effects will be considered in a separate publication.

\section{Conclusions}

In this work we have derived a theory to effectively solve the model of two charged plane-parallel surfaces in the presence of a highly asymmetric electrolyte composed of monovalent salt and in general polyvalent counterions. 
Due to its asymmetric nature we used different approaches to describe different components of the electrolyte solution. The monovalent salt ions were treated on a weak-coupling level and the (polyvalent) counterions on the strong or weak coupling level. This allowed us to formulate a {\em dressed counterion theory} that we solved for various cases: in the WC and SC limits  as well as within the canonical and the grand-canonical ensembles for the dressed counterions. In the dressed counterion theory the salt ions implicitly screen all the interactions in the system turning them from a long range into a short range screened (DH) form.

While salt ions were always treated grand-canonically, counterions were treated either canonically or grand-canonically. The SC dressed counterion theory contains only two free parameters, namely, the salt screening parameter $\kappa$ and the amount of counterions $\eta$ in the slit in the canonical ensemble or $\chi$ (corresponding to the bulk concentration of counterions) in the grand-canonical ensemble. The model itself contains a third free parameter, namely, the coupling parameter $\Xi$ that measures the counterion correlation effects and is considered as $\Xi\to\infty$ in the SC dressed counterion theory. The SC dressed counterion theory is expected to be applicable at finite couplings if the parameter $\Xi$ is high enough. 
In order to test the relevancy of the theory we performed also MC  simulations and compared them with the theory.  

In the canonical case the interaction pressure between the surfaces can become attractive at small to intermediate separations due to strong correlations induced by polyvalent 
counterions. The magnitude of the attraction becomes larger for smaller screening parameters, $\kappa$, and larger amounts of counterions, $\eta$, in the slit. At larger separations the electrostatics is screened out and the counterions behave as an ideal gas. Comparison with MC simulations reveals that our theory is applicable for small enough and large enough separations between the surfaces. As expected, the theory also works better at  larger coupling parameters.

The grand-canonical case, on the other hand,  possesses some qualitative differences compared to the canonical one. At small separations all the counterions are ejected from the inter-surface region and the resulting pressure remains finite in contrast to the diverging contribution of the counterion osmotic pressure in the canonical case. 
The onset of attraction in the grand-canonical ensemble depends on the screening $\kappa$ as well as the bulk concentration of counterions $\chi$.
The MC results confirm that the SC theory is relevant for high enough $\Xi$ and small surface separations as expected from the SC validity criteria.

\section{Acknowledgments}

R.P. would like to acknowledge the financial support by the Slovenian Research Agency under contract
Nr. P1-0055 (Biophysics of Polymers, Membranes, Gels, Colloids and Cells). M.K. would like to acknowledge the financial support by the Slovenian Research Agency under the young researcher grant. 
J.F. acknowledges financial support by the
Swedish Research Council. A.N. is a Newton International Fellow. 

\appendix

\section{Simulations}
\label{app:sim}

Simulations of the model system were performed using the standard Metropolis Monte-Carlo scheme \cite{Metropolis:53}, appropriately adjusted in the case of a grand-canonical system \cite{Norman:69}, when so required for the dressed counterions. The salt ions are treated implicitly using   
the DH pair potential (\ref{DH}) between all other charges. 

In the grand-canonical case, bulk densities were established by separate simulations of a bulk system. In the slit geometry, periodic boundary conditions were applied in the $x-y$ directions parallel to the charged surfaces. The size of the simulation box in these directions, $L$, was always chosen to exceed twice the maximum separation. Furthermore, in all cases we assume $L \gg \kappa^{-1}$ (in most cases, $L$ was considerably larger than $20\kappa^{-1}$). We nevertheless included a long-range interaction correction potential,$\Phi_{LR}(z)$, based upon the average counterion distribution in the slit, $\langle\rho_0(z)\rangle$ \cite{Torrie:82}. Specifically, the rescaled interaction energy between a counterion at a position $z$ and the external (mean-field) distribution is $q\Phi_{LR}(z)$, where 
\begin{equation}
\Phi_{LR}(z)= 2\pi q\ell_{\mathrm{B}}\kappa^{-1}\int_{-a}^a \langle\rho_0(z')\rangle \rme^{-\kappa\sqrt{L^2+(z-z')^2}} {\mathrm{d}}z'.
\end{equation}
Given the typically large size of the simulation box, such a long-range correction is most likely redundant, but its implementation on the
other hand is computationally very cheap. The interaction pressure was calculated across the mid-plane as well as on the surfaces.
The former choice leads to slightly better statistical performance for hard surfaces.



\begin{thebibliography}{99}

\bibitem{holm} 
C. Holm, P. Kekicheff, and R. Podgornik (Eds.), {\em Electrostatic Effects in Soft Matter and Biophysics}, Kluwer Academic, Dordrecht, 2001.

\bibitem{Andelman} 
W. C. K. Poon and D. Andelman (Eds.), {\em Soft condensed matter physics in molecular and cell biology}, Taylor \& Francis, New York, London, 2006.

\bibitem{hoda} 
H. Boroudjerdi, Y. W. Kim, A. Naji, R. R. Netz, X. Schlagberger, and A. Serr,  Phys. Rep. {\bf 416}, 129 (2005).

\bibitem{Naji} 
A. Naji, S. Jungblut, A. G. Moreira, and R. R. Netz, Physica A {\bf 352}, 131 (2005).

\bibitem{shklovskii} 
A. Y. Grosberg, T. T. Nguyen, and B. I. Shklovskii, Rev. Mod. Phys. {\bf 74}, 
329 (2002).

\bibitem{Levin} 
Y. Levin,  Rep. Prog. Phys. {\bf 65}, 1577 (2002).

\bibitem{Rouzina} 
I. Rouzina and V. A. Bloomfield, J. Phys. Chem. {\bf 100}, 9977 (1996).

\bibitem{Netz} 
R. R. Netz, Euro. Phys. J. E {\bf 5}, 557 (2001); A. G. Moreira and R. R. Netz, {\em ibid} {\bf 8}, 33 (2002). 

\bibitem{messina}
R. Messina, J. Phys.: Condens.  Matter {\bf 21}, 113102 (2009). 

\bibitem{podgornik2} 
R. Podgornik, J. Chem. Phys. {\bf 91}, 5840 (1989).

\bibitem{attard} 
P. Attard, J. Mitchell, and B. W. Ninham, J. Chem. Phys. {\bf 88}, 4987 (1988).

\bibitem{podgornik} R. Podgornik and B. \v Zek\v s, J. Chem. Soc., Faraday Trans 2 {\bf 5}, 611 (1988); R. Podgornik, J. Phys. A {\bf 23}, 275 (1990).

\bibitem{fluctuations}
B.-Y. Ha and A. J. Liu, Phys. Rev. Lett. {\bf 79},  1289 (1997); P. A. Pincus and S.A. Safran, Europhys. Lett. {\bf 42}, 103  (1998); M. Kardar and R. Golestanian, Rev. Mod. Phys. {\bf 71}, 1233 (1999); R. R. Netz and H. Orland, Eur. Phys. J. E {\bf 1},  203 (2000); B.-Y. Ha, Phys. Rev. E {\bf 64}, 031507 (2001);  A. W. C. Lau and  P. Pincus, Phys. Rev. E {\bf 66}, 041501  (2002). 

\bibitem{david-ron}
D. S. Dean and R. R. Horgan, J. Phys. C. {\bf 17}, 3473, (2005); Phys. Rev. E {\bf 69}, 061603 (2004); {\em ibid} {\bf 70}, 011101 (2004);   {\em ibid} {\bf 68}, 061106 (2003).

\bibitem{asim} 
M. Kandu\v c, M. Trulsson, A. Naji, Y. Burak, J. Forsman, and R. Podgornik, Phys. Rev. E {\bf 78}, 061105 (2008). 

\bibitem{ziherl}
P. Ziherl, R. Podgornik, and S. \v Zumer, Chem. Phys. Lett. {\bf 295}, 99 (1998). 

\bibitem{podgornikparsegianPRL}
R. Podgornik and V. A. Parsegian, Phys. Rev. Lett. {\bf 80}, 1560 (1997).

\bibitem{original_sims}
L. Guldbrand, B. J\"onsson, H. Wennerstr\"om, and P. Linse,  J. Chem. Phys.
{\bf 80}, 2221 (1984); D. Bratko, B. J\"onsson, and H. Wennerstr\"om,  Chem. Phys. Lett. {\bf 128}, 
449  (1986); J. P. Valleau, R. Ivkov, and G. M. Torrie,  J. Chem. Phys. {\bf 95},  520  (1991); R. Kjellander, T. \AA kesson, B. J\"onsson, and S. Mar\v celja,  J. Chem. Phys. {\bf 97}, 1424  (1992). 

\bibitem{Forsman04}
J. Forsman, J. Phys. Chem. B  {\bf 108}, 9236 (2004). 

\bibitem{intermediate_regime}
Y. Burak, D. Andelman,  and H. Orland,  Phys. Rev. E {\bf 70}, 016102 (2004); 
C. D. Santangelo, Phys. Rev. E {\bf 73}, 041512 (2006); M. M. Hatlo and L. Lue, Soft Matter {\bf 5}, 125 (2009). 

\bibitem{Weeks} 
Y.-G. Chen and J. D. Weeks, Proc. Natl. Acad. Sci. {\bf 103},  7560 (2006); 
J. M. Rodgers, C. Kaur,  Y.-G. Chen, and J. D. Weeks,  Phys. Rev. Lett.
{\bf 97}, 097801 (2006).

\bibitem{arnold}
A. Naji, A. Arnold, C. Holm, and R. R. Netz, Europhys. Lett. {\bf 67}, 130 (2004).

\bibitem{Naji_CCT}
A. Naji and R. R. Netz, Phys. Rev. Lett. {\bf 95}, 185703 (2005); 
Phys. Rev. E {\bf 73}, 056105 (2006).

\bibitem{Jho1}
Y.S. Jho, G. Park, C.S. Chang, P.A. Pincus, and M.W. Kim, 
Phys. Rev. E {\bf 76}, 011920 (2007); {\em ibid} {\bf 73}, 021502 (2006).

\bibitem{trulsson}
M. Trulsson, B. J\"onsson, T. \AA kesson, J. Forsman, and C. Labbez,  Phys. Rev. Lett. {\bf 97}, 068302 (2006); Langmuir {\bf 23}, 11562 (2007).

\bibitem{jho-prl}
Y.S. Jho, M. Kandu{\v{c}}, A. Naji, R. Podgornik, M.W. Kim, and P.A. Pincus, Phys. Rev. Lett. {\bf 101}, 188101 (2008).

\bibitem{exact}
D.S. Dean, R.R. Horgan, and D. Sentenac, J. Stat. Phys. {\bf 90}, 899 (1998);  D.S. Dean, R.R. Horgan, A. Naji, and R. Podgornik,  J. Chem. Phys. {\bf 130}, 094504 (2009). 

\bibitem{Israelachvili} 
J. N. Israelachvili, {\em Intermolecular and Surface Forces}, 2nd edition (Academic Press, London, 1992).

\bibitem{rau-1} 
D.C. Rau and V. A. Parsegian, Biophys J {\bf 61}, 246 (1992).

\bibitem{rau-2} 
D.C. Rau and V. A. Parsegian, Biophys J {\bf 61}, 260 (1992).

\bibitem{olli}
O. Punkkinen, A. Naji, R. Podgornik, I. Vattulainen, and P.-L. Hansen, 
 Europhys. Lett. {\bf 82}, 48001 (2008). 

\bibitem{Edwards} 
S. F. Edwards and A. Lenard, J. Math. Phys. {\bf 3}, 778 (1962).

\bibitem{rudiali}
R. Podgornik and A. Naji, Europhys. Lett. {\bf 74}, 712 (2006). 

\bibitem{kanduc}
M. Kandu{\v c} and R. Podgornik, Eur. Phys. J. E {\bf 23}, 265 (2007). 

\bibitem{PBrepulsive}
J.C. Neu, Phys. Rev. Lett. {\bf 82}, 1072 (1999); J.E. Sader, D.Y.C. Chan, J. Colloid Interface Sci. {\bf 213}, 268 (1999); Langmuir {\bf 16}, 324 (2000); 
E. Trizac, J.-L. Raimbault, Phys. Rev. E {\bf 60}, 6530 (1999); E. Trizac, Phys. Rev. E {\bf 62}, R1465 (2000).

\bibitem{bjerrum} 
N. Bjerrum, K. Dan. Vidensk. Selsk. Mat.-Fys. Medd. {\bf 7}, 1 (1926).

\bibitem{vanroij} 
J. Zwanikken and R. van Roij, J. Phys.: Condens. Matter  {\bf 21}, 424102  (2009).

\bibitem{Fisher} 
M. E. Fisher and Y. Levin, Phys. Rev. Lett. {\bf 71}, 3826 (1993).

\bibitem{Naji-EPJE}
A. Naji and R. R. Netz, Eur. Phys. J. E {\bf 13}, 43 (2004). 

\bibitem{Orr} 
D. Harries, R.Podgornik, V.A. Parsegian, E. Mar-Or, and D. Andelman, J. Chem. Phys. {\bf 124}, 224702 (2006).

\bibitem{note-multipole}
It should be noted that, besides the monopolar charge valency considered here, large counterions may also possess 
charge  quadrupoles and higher-order multipoles due to their internal structure. However, it turns out that, on the SC level, 
multipolar effects can play a role only in a dielectrically inhomogeneous system with dielectric discontinuities at the bounding 
surfaces--see M. Kandu\v c, A. Naji, Y. S. Jho, P. A. Pincus and R. Podgornik, J. Phys.: Condens. Matter  {\bf 21}, 424103  (2009).

\bibitem{Matej-cyl} 
M. Kandu{\v c}, J. Dobnikar, and R. Podgornik, Soft Matter
{\bf 5}, 868 (2009). 

\bibitem{Kjellander} 
R. Kjellander, Colloid J {\bf 69}, 20 (2007). 

\bibitem{alex}
S. Alexander, P. M. Chaikin, P. Grant, G. Morales, P. Pincus, and D. Hone, J. Chem. Phys. {\bf 80}, 5776 (1984).

\bibitem{trizac}
E. Trizac, L. Bocquet, M. Aubouy, and H. von Gr\" unberg, Langmuir {\bf 19}, 4027 (2003).

\bibitem{Metropolis:53} 
N. A. Metropolis and A. W. Rosenbluth and  M. N. Rosenbluth, 
A. Teller and E. Teller, J. Chem. Phys. {\bf 21}, 1087 (1953).

\bibitem{Norman:69} 
G. E. Norman and V. S. Filinov, High Temp. (USSR) {\bf 7} 216  (1969).

\bibitem{Torrie:82}
G. M. Torrie and J. P. Valleau, J. Phys. Chem. {\bf 86}, 3251 (1982).

\end{thebibliography}
\end{document}